\documentclass[prl,aps,twocolumn,superscriptaddress,showpacs,nofootinbib]{revtex4-1}
\usepackage{amscd}
\usepackage{amsthm,amsfonts,dsfont}
\usepackage{enumerate}

\usepackage{mathrsfs}
\usepackage{amssymb}
\usepackage{graphicx}
\usepackage{dcolumn}
\usepackage{bm}
\usepackage{textcomp}
\usepackage{amsmath}
\usepackage[titletoc]{appendix}

\usepackage{epsfig}
\usepackage{epstopdf}
\usepackage{subfigure}
\usepackage{color}
\usepackage[10pt]{moresize}

\newcounter{RomanNumber}

\begin{document}

\title{Twin-field quantum key distribution with local frequency reference}

\author{Jiu-Peng Chen}
\affiliation{Jinan Institute of Quantum Technology and CAS Center for Excellence in Quantum Information and Quantum Physics, University of Science and Technology of China, Jinan 250101, China}
\affiliation{Hefei National Laboratory, University of Science and Technology of China, Hefei 230088, China}

\author{Fei Zhou}
\affiliation{Jinan Institute of Quantum Technology and CAS Center for Excellence in Quantum Information and Quantum Physics, University of Science and Technology of China, Jinan 250101, China}
\affiliation{Hefei National Laboratory, University of Science and Technology of China, Hefei 230088, China}

\author{Chi Zhang}
\affiliation{Jinan Institute of Quantum Technology and CAS Center for Excellence in Quantum Information and Quantum Physics, University of Science and Technology of China, Jinan 250101, China}
\affiliation{Hefei National Laboratory, University of Science and Technology of China, Hefei 230088, China}

\author{Cong Jiang}
\affiliation{Jinan Institute of Quantum Technology and CAS Center for Excellence in Quantum Information and Quantum Physics, University of Science and Technology of China, Jinan 250101, China}
\affiliation{Hefei National Laboratory, University of Science and Technology of China, Hefei 230088, China}

\author{Fa-Xi Chen}
\affiliation{Jinan Institute of Quantum Technology and CAS Center for Excellence in Quantum Information and Quantum Physics, University of Science and Technology of China, Jinan 250101, China}
\affiliation{Hefei National Laboratory, University of Science and Technology of China, Hefei 230088, China}



\author{Jia Huang}
\affiliation{State Key Laboratory of Functional Materials for Informatics, Shanghai Institute of Microsystem and Information Technology, Chinese Academy of Sciences, Shanghai 200050, China}

\author{Hao Li}
\affiliation{State Key Laboratory of Functional Materials for Informatics, Shanghai Institute of Microsystem and Information Technology, Chinese Academy of Sciences, Shanghai 200050, China}

\author{Li-Xing You}
\affiliation{State Key Laboratory of Functional Materials for Informatics, Shanghai Institute of Microsystem and Information Technology, Chinese Academy of Sciences, Shanghai 200050, China}

\author{Xiang-Bin Wang}
\affiliation{Jinan Institute of Quantum Technology and CAS Center for Excellence in Quantum Information and Quantum Physics, University of Science and Technology of China, Jinan 250101, China}
\affiliation{Hefei National Laboratory, University of Science and Technology of China, Hefei 230088, China}
\affiliation{State Key Laboratory of Low Dimensional Quantum Physics, Department of Physics, Tsinghua University, Beijing 100084, China}

\author{Yang Liu}
\affiliation{Jinan Institute of Quantum Technology and CAS Center for Excellence in Quantum Information and Quantum Physics, University of Science and Technology of China, Jinan 250101, China}
\affiliation{Hefei National Laboratory, University of Science and Technology of China, Hefei 230088, China}

\author{Qiang Zhang}
\affiliation{Jinan Institute of Quantum Technology and CAS Center for Excellence in Quantum Information and Quantum Physics, University of Science and Technology of China, Jinan 250101, China}
\affiliation{Hefei National Laboratory, University of Science and Technology of China, Hefei 230088, China}
\affiliation{Hefei National Research Center for Physical Sciences at the Microscale and School of Physical Sciences, University of Science and Technology of China, Hefei 230026, China}

\author{Jian-Wei Pan}
\affiliation{Hefei National Laboratory, University of Science and Technology of China, Hefei 230088, China}
\affiliation{Hefei National Research Center for Physical Sciences at the Microscale and School of Physical Sciences, University of Science and Technology of China, Hefei 230026, China}

\begin{abstract}
Twin-field quantum key distribution (TF-QKD) overcomes the linear rate-loss limit, which promises a boost of secure key rate over long distance. However, the complexity of eliminating the frequency differences between the independent laser sources hinders its practical application. Here, taking the saturated absorption spectroscopy of acetylene as an absolute reference, we propose and demonstrate a simple and practical approach to realize TF-QKD without requiring relative frequency control of the independent laser sources. Adopting the 4-intensity sending-or-not-sending TF-QKD protocol, we experimentally demonstrate the TF-QKD over 502 km, 301 km and 201 km ultra-low loss optical fiber respectively. We expect this high-performance scheme will find widespread usage in future intercity and free-space quantum communication networks.
\end{abstract}
\maketitle

{\it Introduction.---}
Quantum key distribution (QKD)~\cite{bennett1984quantum,gisin2002quantum,gisin2007quantum,scarani2009security,xu2020secure,pirandola2020advances} offers an information-theoretically secure way to share secure keys between distant users. Since the quantum signal is forbidden to be amplified~\cite{1982Wootters} and decays exponentially with the transmission distance, without a quantum repeater, the point-to-point secret key capacity (SKC) scales linearly with the channel transmission~\cite{PLOB2017}, which poses an inevitable barrier for long distance QKD. As an efficient version of measurement-device-independent QKD~\cite{lo2012measurement,braunstein2012side}, the twin-field QKD (TF-QKD)~\cite{lucamarini2018overcoming} improves the secure key rate to overcome the linear rate-loss limit, which enhances the key rate to the square root scale of the channel transmittance with current available technologies. Therefore, the combination of measurement-device-independence and excellent tolerance on channel loss made TF-QKD rapidly becoming the focus of competing research once it was proposed. At present, TF-QKD has been obtained many achievements in theory~\cite{lucamarini2018overcoming,PhysRevX.8.031043,wang2018sns,Curty2019,PhysRevA.100.022306} and experiment~\cite{minder2019experimental,wang2019beating,liu2019exp,zhong2019proof,chen2020sending,fang2019surpassing,liu2021field,pittaluga2021600,chen2021twin,Wang2022,PhysRevLett.128.180502,Zhou2023,PhysRevLett.130.250801,PhysRevLett.130.250802,PhysRevLett.130.210801}.
These efforts pave the way for the realization of long-distance quantum communication networks with enhanced security and improved performance.

However, implementing TF-QKD is challenging, because the protocols require coherently controlling the twin light fields from remote parties. Any phase differences, caused by frequency differences between independent lasers or channel fiber fluctuations may disturb the coherence of the twin light fields. Currently, the phase differences caused by hundreds of kilometers of fiber fluctuations are generally limited and can be effectively compensated using mature techniques, either in real-time~\cite{wang2019beating,pittaluga2021600,Wang2022,PhysRevLett.130.250801} or through post-processing methods~\cite{liu2019exp,fang2019surpassing,chen2020sending,liu2021field,chen2021twin,PhysRevLett.128.180502,PhysRevLett.130.210801}. The fast phase variations originating from the light sources without frequency locking can be much more severe than that caused by long fiber fluctuations. By fast frequency locking such as time-frequency metrology~\cite{liu2019exp,chen2020sending,chen2021twin,PhysRevLett.128.180502,PhysRevLett.130.250801,Zhou2023,PhysRevLett.130.210801}, or the optical phase locking loop (OPLL)~\cite{wang2019beating,minder2019experimental,pittaluga2021600,Wang2022}, the relative frequency differences are real-time eliminated with a gigantic and complicated settings on light sources. Alternatively, through the high-speed single photon detection and the fast Fourier transform (FFT) algorithm~\cite{PhysRevLett.130.250802}, the relative frequency differences could be eliminated in post-processing with a high count measurement devices and complex data post-processing operations in measurement station. All these previous methods could potentially hinder its wide application.

Moreover, the scarcity of free-space link channels means that an additional channel for frequency locking~\cite{wang2019beating,minder2019experimental,liu2019exp,chen2020sending,pittaluga2021600,chen2021twin,Wang2022,PhysRevLett.128.180502,PhysRevLett.130.250801,Zhou2023,PhysRevLett.130.210801} could increase the cost and complexity of free-space TF-QKD. Meanwhile, generating secure keys on small data sizes is inevitable in free-space experiments due to their weather-dependence, therefore data post-processing operations~\cite{PhysRevLett.130.250802} could impede efficient data collection during implementation of TF-QKD in free-space.

Here, applying two acetylene cells as the absolute frequency standard to eliminate the frequency drift of Alice's and Bob's seed fiber lasers, the frequency differences of the two light sources are restricted to vary in a small range of less than 300 Hz. Without relative frequency control of independent laser sources, we experimentally demonstrated 4-intensity sending-or-not-sending (SNS) TF-QKD~\cite{wang2018sns} with the actively-odd-parity-pairing (AOPP)~\cite{jiang2019unconditional,xu2019general,jiang2021composable,Hu_2022} method over 502 km, 301 km and 201 km ultra-low loss (ULL) optical fiber links respectively. To get high key rates, we apply the advanced decoy-state analysis method by putting the error correction process before the decoy-state analysis which allows us to take all vacuum-related counts in the decoy-state analysis, and the advanced key distillation scheme by taking the counts associated with the strongest two light sources as the raw keys to extract the final keys (See Supplemental Material for details of the protocol.). Furthermore, without servo-induced noise on rapidly frequency locking on laser sources, a high performance single photon interference to support a low phase flip error rate of less than 3\% is obtained, thus the final secure key rates corresponding to a total sent pulses as few as $2\times 10^{11}$ are still considerable. In other words, if the similar system frequency of about 1~GHz to that in Ref.~\cite{PhysRevLett.130.210801} and Ref.~\cite{Wang2022} is used, a considerable secure key rate can be obtained at the minutes level.

In the promotion of the practical application of quantum communication, which overcomes the linear rate-loss limit, there are two related works~\cite{PhysRevLett.130.030801,PhysRevLett.130.250801} that demonstrate different approaches by applying the so-called mode-pairing protocol. Without relative frequency locking between independent laser sources, Ref.~\cite{PhysRevLett.130.030801} and Ref.~\cite{PhysRevLett.130.250801} simplify the setup compared to previous demonstrations on TF-QKD. Nonetheless, the final secure key rate of the mode-pairing protocol depends entirely on the post-selecting time slots within the coherence time of the laser sources. As a result, without ultra-stable laser sources that have a much longer coherence time but add more complexity and cost~\cite{PhysRevLett.130.250801}, the final secure key rate and tolerable transmission loss are lower than those of TF-QKD.

The experimental setup is shown in Fig~\ref{Fig:seisim-setup}(a). Alice and Bob use two continuous wave (CW) fiber lasers with a linewidth of several hundreds Hz which are referenced to the saturated absorption spectroscopy of acetylene~\cite{Hald:11,Balling:05} as their light sources. The  narrow linewidth CW light beam with a central wavelength of 1542.3837 nm is then modulated to a waveform pattern (as shown in Fig~\ref{Fig:seisim-setup}(b)) that the single-photon-level quantum signal pulses are time multiplexed with strong phase reference pulses. The signals from Alice and Bob are sent to Charlie through ULL fiber spools with an average transmission loss of about~$0.167~dB/km$. After interference at Charlie's beam splitter (BS), the signals are detected by two superconducting nanowire single photon detectors (SNSPDs), and recorded by a time tagger.

\begin{figure*}[htb]
	\centering
	\resizebox{14cm}{!}{\includegraphics{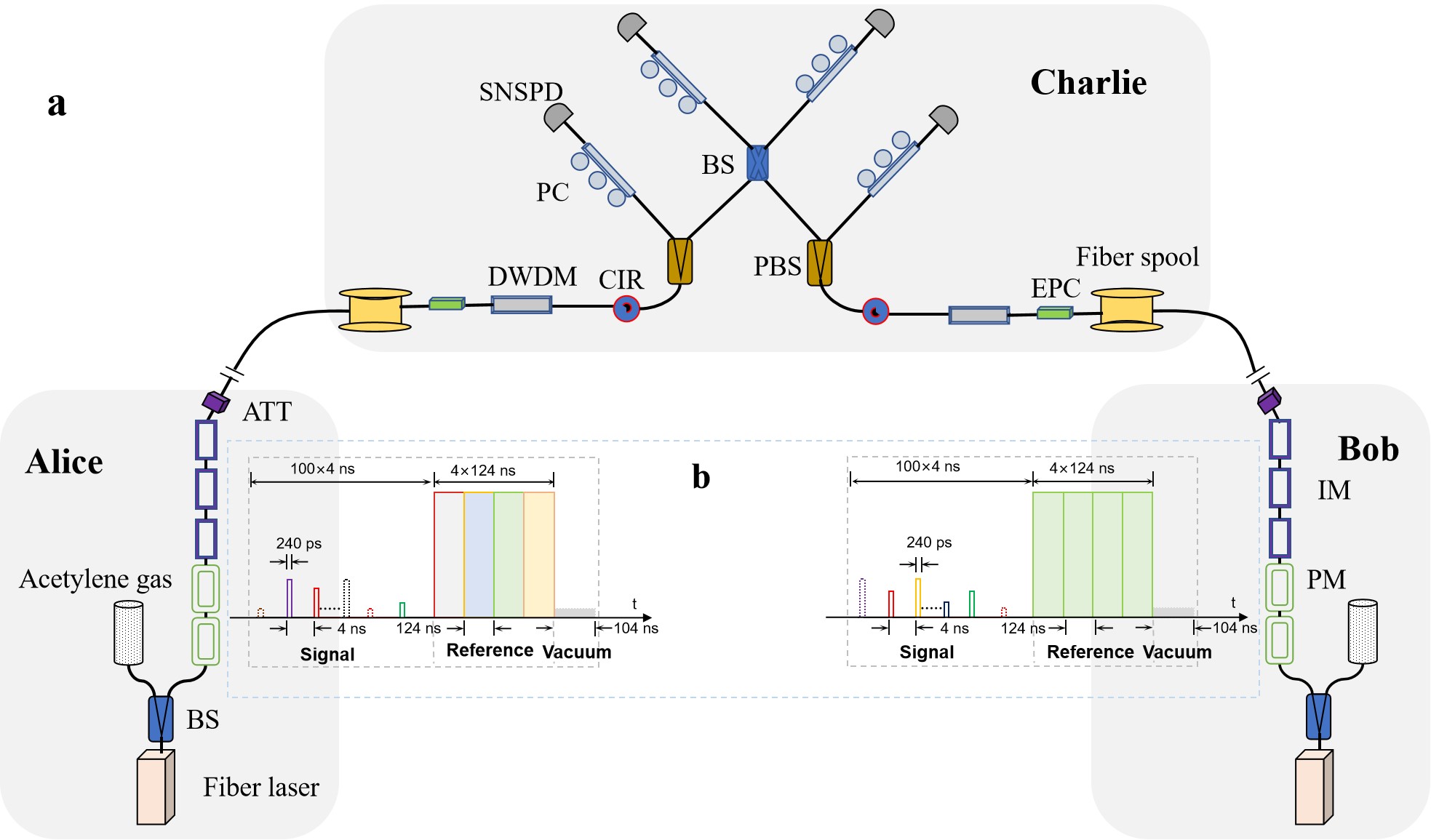}}
	\caption{(a) Schematic of experimental setup. \baselineskip12ptIn Alice's and Bob's labs, two fiber seed lasers with a linewidth of about 1~kHz are referenced to the saturated absorption spectroscopy of an acetylene cell individually. With a central wavelength of 1542.3837 nm, the frequency differences between the two lasers are restricted to slowly fluctuating within a range of 300~$Hz$. Then the acetylene-stabilized light sources are modulated with 2 phase modulators (PMs) and 3 intensity modulators (IMs) to a waveform pattern that time multiplexes the weak signal pulses with strong phase reference pulses, and  attenuates them to bring the signal pulses to the single-photon level with an attenuator (ATT). The prepared light pulses are finally sent to Charlie through the ultra-low loss fiber spools for detection. Charlie uses a Dense Wavelength Division Multiplexer (DWDM), and a circulator (CIR) to filter the noises before the polarization beam splitter (PBS) and the beam splitter (BS). The interference results are detected by superconducting nanowire single-photon detectors (SNSPDs). EPC: electric polarization controller, PC: polarization controller. (b) Waveform pattern of the modulation. \baselineskip12ptThe CW light beam is modulated to a waveform pattern that 100 signal pulses are time multiplexed with 4 strong phase reference pulses in a basic period. For each basic period of 1 $\mu$s time sequence, 100 signal pulses with 4 random intensities and 16 random phase values are prepared in the first 400 ns, each with a 240~ps pulse duration and 3.76~ns interval. Then 4 strong phase reference pulses with the same intensity and fixed phase values are prepared in the following 496~ns, each with a 124~ns pulse duration. Finally, a 104-ns extinction pulse is prepared as the recovery time for the SNSPDs.}
	\label{Fig:seisim-setup}
\end{figure*}

Crucially, the realization of TF-QKD involves controlling the phase evolution of the twin fields, which travel hundreds of kilometers through the channel before interfering at Charlie's BS. The differential phase fluctuation between the two optical fields sent from users hundreds of kilometers apart to Charlie can be written as~\cite{lucamarini2018overcoming}:
\begin{equation}\label{eq:phase}
\begin{split}
\delta_{\mathrm{ba}}=\frac{2 \pi}{s}(\Delta \nu L+\nu \Delta L)
\end{split}
\end{equation}

The first term in Eq.~\eqref{eq:phase} represents the frequency difference ($\Delta \nu$) between the users' lasers, while the second term denotes the fluctuation of the long fiber paths. In previous TF-QKD demonstrations, excessive resources were dedicated to eliminating the relative frequency differences between laser sources, i.e., the first term in Eq.~\eqref{eq:phase}.

We solve this problem by setting two acetylene cells in Alice's and Bob's labs, and eliminating the frequency drift of the light sources referring to their saturated absorption spectroscopy with a long stability of about $3\times 10^{-13}$ respectively (See Supplemental Material for details of the acetylene-stabilized laser.). The specific transitions in acetylene molecules have well-defined and precise frequencies, making them suitable for utilizing in optical frequency standards to serve as a highly stable frequency reference~\cite{Hald:11,Balling:05}. With the help of the two acetylene cells, the frequency differences between the two independent fiber lasers are restricted to slow variation within a small range of 300~Hz. As shown in Fig~\ref{Fig:frequency} (a), the frequency differences variation of the two acetylene-stabilized light sources are less than 450~Hz in 40 minutes with a standard deviation of about 60~Hz. Under the influence of the frequency differences, as shown in Fig~\ref{Fig:frequency} (b), the phase fluctuation rate of the two independent light sources is~$0.015~rad/us$, which is comparable to the fluctuation of hundreds of kilometers of fiber links in the field~\cite{chen2021twin} and can be eliminated through fiber fluctuation suppression. 

\begin{figure}[htb]
	\centering
	\resizebox{8cm}{!}
	{\includegraphics{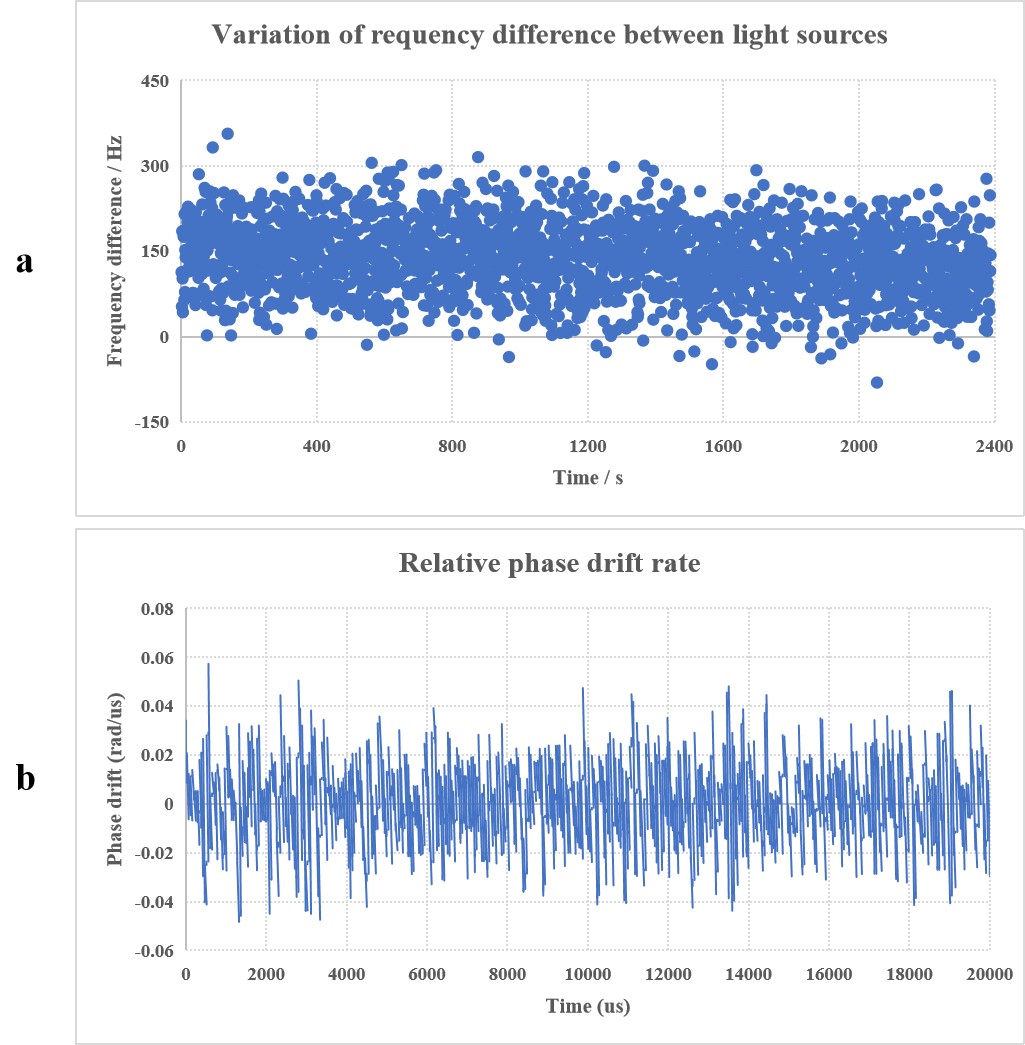}}
	\caption{\baselineskip12pt(a) Variation of the frequency difference between the light sources. (b) Relative phase drift rate between the light sources.}
	\label{Fig:frequency}
\end{figure}

After setting the light sources, two phase modulators (PM) and three intensity modulators (IM) are inserted to modulate the CW light beam into a waveform pattern that 100 signal pulses are time multiplexed with 4 strong phase reference pulses in a basic period. For each basic period of 1~$\mu$s time sequence, 100 signal pulses with 4 random intensities and 16 random phase values are prepared in the first 400~ns, each with a 240~ps pulse duration and 3.76~ns interval. Then 4 strong phase reference pulses with the same intensity and fixed phase values are prepared in the following 496~ns, each with a 124~ns pulse duration. Finally, a 104-ns extinction pulse is prepared as the recovery time for the SNSPDs. Before the prepared pulses are sent out of Alice's and Bob's labs, they are attenuated on both sides to bring the signal pulses to the single-photon level with passive attenuators. Through two symmetrical fiber links consisting of ultra low loss fiber spools, the two pulse trains arrive at Charlie, and interfere at a BS. The interference results are detected by 2 SNSPDs and recorded by a high-speed multichannel time tagger.

\begin{figure}[htb]
	\centering
	\resizebox{8cm}{!}
	{\includegraphics{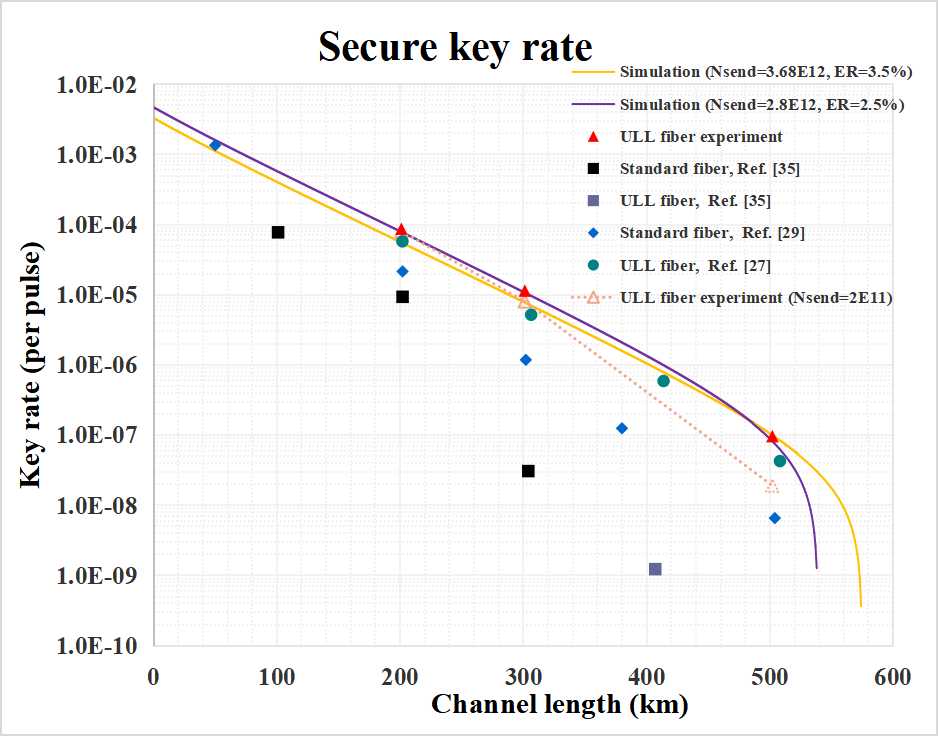}}
	\caption{\baselineskip12ptSecure key rates of the SNS-TF-QKD experiment. The red triangles indicate the experimental results over 201 km, 301 km ULL fiber with a total sent pulses of $2.87\times 10^{12}$, and 502 km ULL fiber with a total sent pulses of $3.68\times 10^{12}$, the brown triangles indicate the experimental results while the total sent pulses are curtail to a small size of $2\times 10^{11}$. The green dot, and purple square indicate the experimental results of ref.~\cite{PhysRevLett.130.250801} and ref.~\cite{PhysRevLett.130.030801} over the ULL fiber. The black square and blue diamond indicate the experimental results of ref.~\cite{PhysRevLett.130.030801} and ref.~\cite{PhysRevLett.130.250802} over the standard fiber. The purple curve is the simulation result with our experimental parameters that a total of $2.87\times 10^{12}$ pulses are sent and the phase flip error rate is 2.5\%. The orange curve shows the simulation result with our experimental parameters that a total of $3.68\times 10^{12}$ pulses are sent and the phase flip error rate is 3.5\%.}
	\label{Fig:key-rate}
\end{figure}

To suppress system noise, before the twin field light pulses from Alice and Bob interfere on the BS in Charlie, a dense wavelength division multiplexer (DWDM) at the central wavelength of 1548.15~nm with a bandwidth of 100~GHz is inserted to filter the nonlinear scattering light which originates from the strong phase reference pulses~\cite{chen2020sending}. Following a circulator is inserted to prevent the strong reference pulses from being reflected off the end face of the SNSPDs into the optical fiber links to scatter backward noise~\cite{chen2020sending}.

To maintain identical polarization and arrival time of the twin field light pulses at Charlie's BS, feedback devices for channel delay and polarization of signal pulses are incorporated into the experimental setup. Before the twin field light pulses from Alice and Bob interfere on Charlie's BS, a polarization beam splitter (PBS) is inserted. By monitoring the idle beam of the PBS in real-time to detect the count rate and rise time of the first strong phase reference pulse, the variation of the polarization and channel delay caused by fiber paths fluctuation can be obtained. Then, an electric polarization controller (EPC) is inserted to calibrate the polarization. Meanwhile, by adjusting the clock phase of the encoding signal sources about every 20 seconds, the variation of the channel delay is compensated.

Along with the perturbation of polarization and arrival time, the fiber paths fluctuation an also cause disturbances in the globe phase of the signal pulses, which can be eliminated through post-data selection method~\cite{liu2019exp}. The detections of the interference between the strong phase reference pulses are controlled to not below 1.5 MHz per channel, and the relative phase difference is estimated for every 10~$\mu$s with a count of about 30 and compensated in post-data selection.

After considering all of the above, we performed symmetrical 4-intensity SNS-TF-QKD over a total length of 201 km, 301 km, and 502 km of ULL optical fibers, respectively. The corresponding total losses are 33.6 dB, 50.4 dB, and 83.7 dB, including the connections, which average to 0.167 dB/km. The total insertion loss of the optical components is optimized to 1.8 dB in Charlie. Next, we adopted two high-performance SNSPDs with a detection efficiency of 70\% and 72\%, along with an effective dark count rate of 0.2 Hz for both, to detect the interference results. We set a time gate of 0.3 ns to suppress noise, resulting in an additional loss of 1.2 dB. Taking into account the finite data size effect~\cite{jiang2019unconditional,jiang2020zigzag}, we then calculate the secure key rate~\cite{Hu_2022}. The error correction efficiency is $f = 1.16$. The failure probabilities of the error correction and privacy amplification processes, as well as the application of the Chernoff bound in finite-size estimation, are all set to $1\times 10^{-10}$.

For the demonstrations on the lengths of 201 km and 301 km optical fibers, a total of $2.87\times 10^{12}$ signal pulses were sent at an effective system frequency of 100 MHz, resulting in $4.33\times10^9$ and $6.75\times10^8$ valid detections, respectively. Following this, a total of $3.68\times 10^{12}$ signal pulses were sent on the length of 502 km optical fiber, with $1.90\times10^7$ valid detections. We observed a homologous quantum phase flip error rate (QBER) in X basis of around 3.0\%, 3.1\%, and 4.5\% with a base-line error rate of around 2.8\%. The bit-flip error rate in Z basis was 22.23\%, 21.97\%, and 25.05\% before AOPP and decreased to 0.00246\%, 0.0103\%, and 0.392\% after AOPP, while the phase error rate increased to 7.8\%, 8.39\%, and 13.32\%. We then summarized our theoretical simulation and experimental results in Fig.~\ref{Fig:key-rate}. The obtained secure key rates here were $R=8.74\times10^{-5}$, $R=1.15\times10^{-5}$, and $R=9.67\times10^{-8}$, respectively. Furthermore, even when the total number of sent pulses was reduced to $2\times 10^{11}$, the corresponding secure key rates still remained considerable at $R=1.74\times10^{-5}$, $R=2.68\times10^{-6}$, and $R=2.57\times10^{-8}$ (See Supplemental Material for details of the experimental results.).

In conclusion, we proposed and demonstrated a practical approach to realize TF-QKD using local optical oscillators by refereing to an absolute frequency standard, the saturation absorption of acetylene, without requiring relative frequency control of independent laser sources. With our setup, adopting the 4-intensity SNS-TF-QKD protocol with the AOPP method, we experimentally demonstrated TF-QKD over ULL fibers of lengths 502 km, 301 km, and 201 km, respectively. Our work simplifies and provides an effective and practical solution to TF-QKD, taking an important step towards a wide range of applications, especially for realizing free-space TF-QKD where channel resources are scarce, and continuity of implementation is weather-dependent.


{\it Acknowledgments.---}


This work was supported by the National Key Research and Development (R\&D) Plan of China (Grant No. 2020YFA0309800), 
the Innovation Program for Quantum Science and Technology (2021ZD0300700), 
the National Natural Science Foundation of China (Grant Nos. T2125010, 
12174215, 
61971409 and 61971408), 
the Chinese Academy of Sciences, the Key R\&D Plan of Shandong Province (Grant No. 2023CXPT105, 2020CXGC010105), Shandong provincial natural science foundation (Grant Nos. ZR2022LLZ011, ZR2020YQ45 
). 
Y.L. and Q.Z. acknowledge support from the Taishan Scholar Program of Shandong Province.

Jiu-Peng Chen, Fei Zhou and Chi Zhang contributed equally to this work.

\maketitle
\section{Appendix}
\section*{Protocol}

The 4-intensity SNS-TF-QKD with the actively-odd-parity-pairing (AOPP) method is applied in this work. To get high key rates, we also apply the advanced decoy-state analysis method by putting the error correction process before the decoy-state analysis which allows us takes all vacuum-related counts in the decoy-state analysis, and the advanced key distillation scheme by taking the counts associated with the strongest two light sources as the raw keys to extract the final keys~\cite{Hu_2022}.

In the 4-intensity SNS-TF-QKD, there are four sources in Alice's (Bob's) side which are the vacuum source $v$, the weak coherent state sources $x,y,z$ with intensities $\mu_v=0,\mu_x,\mu_y,\mu_z$ and probabilities $p_0,p_x,p_y,p_z$ respectively. In each time window, Alice (Bob) randomly prepares and sends out a pulse from the four candidate sources to Charlie who is assumed to measure the interference result of the incoming pulse pair and announce the measurement results to Alice and Bob. If only one of the two detector clicks, Alice and Bob would take it as an effective event.  After Alice and Bob send $N$ pulse pairs to Charlie, and Charlie announces all measurement results, Alice and Bob perform the following data post-processing.

We denote $lr$ as the two-pulse source while Alice chooses the source $l$ and Bob choose the source $r$ for $l,r=v,x,y,z$. We denote the number of effective events from source $lr$ as $n_{lr}$. In the data post-processing, Alice and Bob first announces the location of the time windows that either Alice or Bob chooses the source $x$. After this, Alice and Bob shall know the location of the time windows while they use the sources $vx,xv,xx$ etc., while keep the locations of the the time windows while they use the sources $vv,vy,yv,vz,zv,yz,zy,yy,zz$ secret (un-announced). For those unannounced times windows, Alice (Bob) takes it as bit $0$ ($1$) if she (he) chooses the source $v$, and takes it as bit $1$ ($0$) if she (he) chooses the source $y$ or $z$, and the bits of those corresponding effective event from the un-announced windows form the raw key strings $Z_A$ and $Z_B$ in Alice and Bob's sides respectively. Alice and Bob first perform the AOPP process to the raw key strings and then perform the error correction process. After this, Alice shall know the value of $n_{yv}$, Bob shall know the value of $n_{vy}$ and both of them shall know the value of $n_{vv}$. Then Alice announces the value of $n_{yv}$ to Bob. With all those values of $n_{vv},n_{vx},n_{xv},n_{vy},n_{yv}$, Bob can perform the decoy state analysis to get the lower bound of the number of untagged bits after AOPP. Also, according to the phase information of the effective events from source $xx$, Bob can get the upper bound of phase flip error rate of untagged bits after AOPP. Finally, Alice and Bob can calculate  the secure key rate according to the formulas shown in Ref.~\cite{Hu_2022}.

\section*{Acetylene-stabilized laser sources}
\begin{figure}[htb]
	\centering
	\resizebox{8cm}{!}
	{\includegraphics{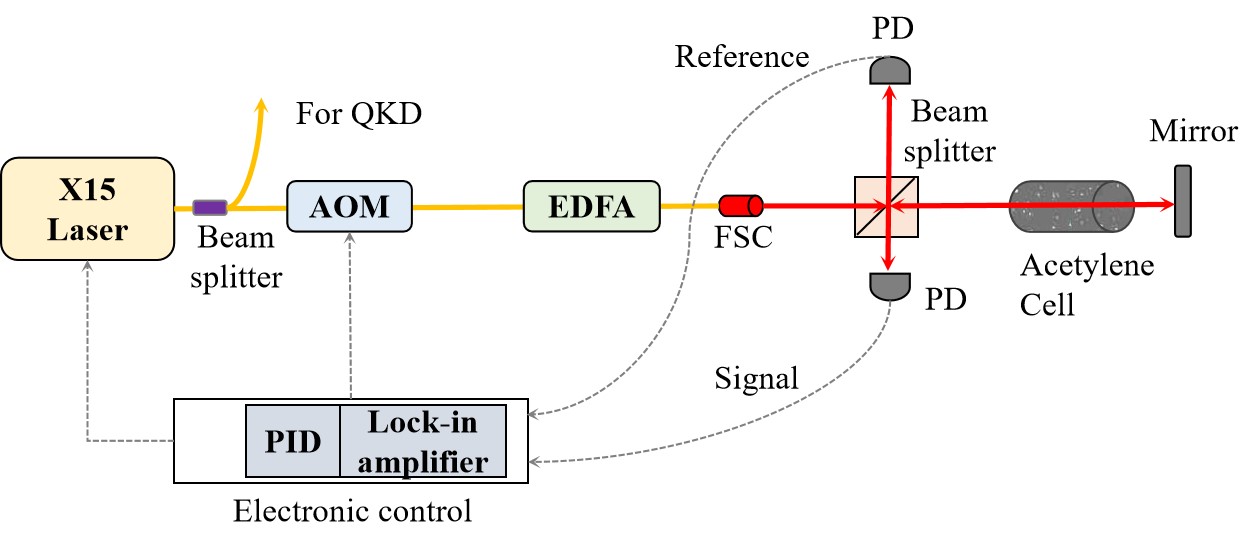}}
	\caption{\baselineskip12ptSchematic of the STABILASER 1542$^\varepsilon$ module. AOM: acousto-optic modulator, PD: photo detector, FSC: fiber-to-free space coupler, EDFA: erbium-doped fiber amplifier.}
	\label{Fig:laser}
\end{figure}

The acetylene-stabilized laser is the STABILASER 1542$^\varepsilon$ module~\cite{Hald:11,Balling:05} from Danish National Metrology Institute, which design is sketched in Fig~\ref{Fig:laser}. The seed fiber laser of Koheras BASIK X15 module from NKT Photonics is temperature tuned to the saturated absorption line ${ }^{13} \mathrm{C}_{2} \mathrm{H}_{2} \mathrm{P}(16)\left(v_{1}+v_{3}\right)$ of acetylene cell at 1542.3837 nm . A piezo tuning integrated in X15 is used for fast fine-tuning of its output frequency with a range of 3 GHz. An acousto-optic modulator (AOM) is used to frequency shift and frequency modulate part of the output beam of the X15 with a shift of about 40 MHz, a modulation frequency of about 1 kHz, and a peak-to-peak frequency modulation width of about 1 MHz. Then pass through an Erbium Doped Fiber Amplifier (EDFA) before free space coupling into the acetylene cell. A small fraction of the beam before and after the passage through the gas cell is reflected off the beam-splitter and provides the reference and signal beams for the balanced detection. Following, a lock-in amplifier is applied to detect the third harmonic content in the photoreceiver output with a narrow bandwidth to improve the signal-to-noise ratio. The third harmonic signal is integrated in a PID (Proportional-Integral-Derivative) controller. The PID output is amplified in the piezo driver and then added to the modulation signal entering the laser piezo input.

\begin{figure}[htb]
	\centering
	\resizebox{8cm}{!}
	{\includegraphics{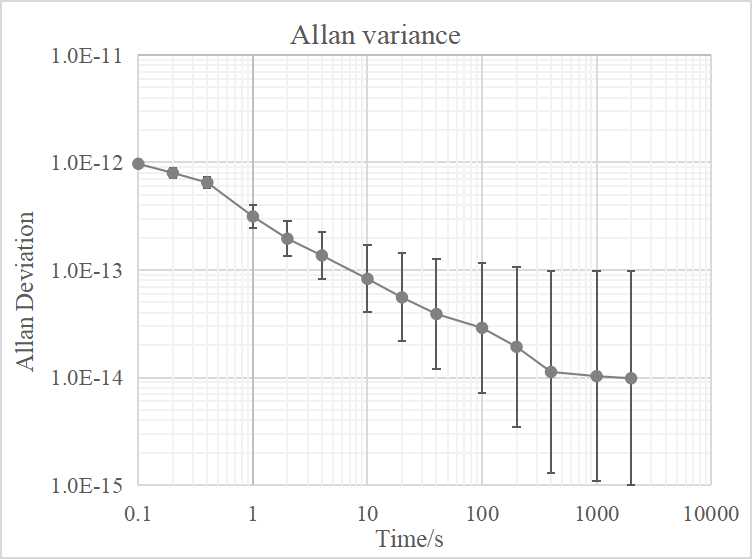}}
	\caption{\baselineskip12ptAllan variance of the two lase sources.}
	\label{Fig:Allan}
\end{figure}

The relative Allan variance of the beat note between the two acetylene-stabilized laser sources is presented in Fig~\ref{Fig:Allan}, which is below $3\times 10^{-13}$ for integration times above 1~s.

\section*{Detailed Experimental Results}
We denote $lr$ as the two-pulse source while Alice chooses the source $l$ and Bob choose the source $r$ for $l,r=v,x,y,z$. Tab.~\ref{Tab:IntensityProbability} lists the intensities and probabilities of different chosen sources. $\mu_v$, $\mu_x$, $\mu_y$ and $\mu_z$ are the choice intensities for the source of $v,x,y,z$ respectively. $p_v$, $p_x$, $p_y$ and $p_z$ are the choice probabilities of $v,x,y,z$.

\begin{table*}[htb]
\centering
  \caption{Intensities and sending ratios of different states.}
\begin{tabular}{c|c|c|c}
\hline
Total channel length &201 km & 301 km& 502 km\\
\hline
$\mu_x$	&      0.092 & 0.100 &     0.100\\
$\mu_y$	&      0.342&  0.345 &    0.387\\
$\mu_z$	&      0.463&  0.459 &    0.470\\
\hline
$p_v$	&      0.700&  0.700 &    0.623\\
$p_x$	&      0.050&  0.050 &    0.133\\
$p_y$	&      0.050&  0.050 &    0.044\\
$p_z$	&      0.2&    0.2   & 0.2\\
\hline
\end{tabular}
\label{Tab:IntensityProbability}
\end{table*}

Tab.~\ref{Tab:Result1}, Tab.~\ref{Tab:Result2}, Tab.~\ref{Tab:Result3} summarizes the experimental results. It shows the total sending number of signal pulses ($N_{total}$) and the final key rate $R$ for the best possible accepted phase difference $Ds$ (in degrees). In our experimental implementation, we applied a digital window to select the signal in the middle of each pulse, where the interference is expected to be better and all the detections are filtered according to it, before the detections are announced by Charlie. The results of un-tagged bits ($n_1$), phase-flip error rate $e_1^{ph}$ and the bit error rates before and after applying the AOPP method are also given respectively. ``QBER" represents for the error rates of Alice's and Bob's chosen source are the $vv,vy,vz,yv,yy,yz,zv,zy,zz$.

In the following rows, we list the numbers of pulses Alice and Bob sent in different sources, labelled as ``Sent-lr"; ``l" (``r") is ``v", ``x",  ``y" or  ``z", indicating the intensity which Alice (Bob) has chosen within ``vacuum", ``$\mu_x$", ``$\mu_y$" or ``$\mu_z$". As with the numbers of sent pulses, the numbers of detections are listed as ``Detected-lr". The numbers of valid detections before AOPP are listed as ``Detected-Valid-(Before AOPP)'' and the corresponding survived bits after AOPP are listed as ``Survived-bits(After AOPP)''. The total valid detections reported by Charlie is denoted as ``Detected-Valid-ch", where ``ch" can be ``Det1" or ``Det2" indicating the responsive detector of the recorded counts. The table also gives the numbers of detections falling within the accepted difference range $Ds$, listed as ``Detected-lr-Ds-Ch", where ``Ds" indicates that only the data within the accepted range $Ds$ are counted, ``Ch" indicates the detection channel.

\linespread{1.5}
\begin{table*}[htb]
\centering
  \caption{Experimental results.}
\begin{tabular}{c|c|c|c}
\hline
\hline
Fiber Length & 201 km &Fiber loss &33.6~dB\\
\hline
$Ds$		 & $8^\circ$  &$R$    & $8.74\times 10^{-5}$ \\
\hline
$N_{total}$	 & $2.87\times10^{12}$  &Detected-total & 4332682589	\\
Sent-vv	 & 1404080000000   &Detected-vv	& 49154	  	\\
Sent-vx	 & 101535000000	   &Detected-vx	& 63710095	\\
Sent-vy	 & 101536000000    &Detected-vy	& 232388800	\\
Sent-vz	 & 401792000000	   &Detected-vz	& 1260521661	\\
Sent-xv	 & 101535000000	   &Detected-xv	& 69370175	\\
Sent-xx	 & 7253000000	   &Detected-xx	& 8875192 \\
Sent-xy	 & 7252000000	   &Detected-xy	& 21744273	\\
Sent-xz	 & 27559000000	   &Detected-xz	& 220656658	\\
Sent-yv	 & 100083000000	   &Detected-yv	& 220656658	\\
Sent-yx	 & 7253000000     &Detected-yx	& 32790166	\\
Sent-yy	 & 7253000000      &Detected-yy	& 32790166 \\
Sent-yz	 & 29010000000     &Detected-yz	& 157091004  \\
Sent-zv	 & 403243000000    &Detected-zv	& 1185649771 \\
Sent-zx	 & 29009000000 	   &Detected-zx	& 103308019 \\
Sent-zy	 & 29000000000     &Detected-zy	& 152136382 \\
Sent-zz	 & 114588000000    &Detected-zz	& 699085188 \\
\hline
$n_1$(Before AOPP)		 & 2058250000  &$n_1$(After AOPP) & 415739000 \\
$e_1^{ph}$(Before AOPP)  & 4.031\% &$e_1^{ph}$(After AOPP)   & 7.802\%\\
QBER$(Before AOPP)$ 	 & 26.423\%  &QBER$(After AOPP)$ 	 & 0.002\%  \\
Detected-Valid(Before AOPP) &4332682589 &Survived-bits(After AOPP)& 828544000\\
\hline
\end{tabular}
\label{Tab:Result1}
\end{table*}

\linespread{1.5}
\begin{table*}[htb]
\centering
  \caption{Experimental results.}
\begin{tabular}{c|c|c|c}
\hline
\hline
Fiber Length & 301 km &Fiber loss &50.4~dB\\
\hline
$Ds$		 & $8^\circ$  &$R$    & $1.15\times 10^{-5}$ \\
\hline
$N_{total}$	 & $2.87\times10^{12}$  &Detected-total & 698588178	\\
Sent-vv	 & 1403600000000   &Detected-vv	& 33477	  	\\
Sent-vx	 & 101500000000	   &Detected-vx	& 10160767	\\
Sent-vy	 & 101500000000	   &Detected-vy	& 33953160	\\
Sent-vz	 & 401650000000	   &Detected-vz	& 182277776	\\
Sent-xv	 & 101500000000	   &Detected-xv	& 11551224	\\
Sent-xx	 & 7250000000	   &Detected-xx	& 1525200 \\
Sent-xy	 & 7250000000	   &Detected-xy	& 3256718	\\
Sent-xz	 & 27550000000	   &Detected-xz	& 15677407	\\
Sent-yv	 & 100050000000	   &Detected-yv	& 40083383	\\
Sent-yx	 & 7250000000     &Detected-yx	& 3632420	\\
Sent-yy	 & 7250000000      &Detected-yy	& 5404053  \\
Sent-yz	 & 29000000000     &Detected-yz	& 25498831  \\
Sent-zv	 & 403100000000    &Detected-zv	& 211108166 \\
Sent-zx	 & 29000000000 	   &Detected-zx	& 18458544 \\
Sent-zy	 & 29000000000     &Detected-zy	& 25467525 \\
Sent-zz	 & 114550000000    &Detected-zz	& 110499527 \\
\hline
$n_1$(Before AOPP)		 & 295150000  &$n_1$(After AOPP) & 56744100 \\
$e_1^{ph}$(Before AOPP)  & 4.338\% &$e_1^{ph}$(After AOPP)   & 8.393\%\\
QBER$(Before AOPP)$ 	 & 26.312\%  &QBER$(After AOPP)$ 	 & 0.010\%  \\
Detected-Valid(Before AOPP) &698588178 &Survived-bits(After AOPP)& 141810000\\
\hline
\end{tabular}
\label{Tab:Result2}
\end{table*}

\linespread{1.5}
\begin{table*}[htb]
\centering
  \caption{Experimental results.}
\begin{tabular}{c|c|c|c}
\hline
\hline
Fiber Length & 502 km &Fiber loss &83.7~dB\\
\hline
$Ds$		 & $10^\circ$  &$R$    & $9.67\times 10^{-8}$ \\
\hline
$N_{total}$	 & $3.68\times10^{12}$  &Detected-total & 18973852	\\
Sent-vv	 & 1421668000000   &Detected-vv	& 25403	  	\\
Sent-vx	 & 308576000000	   &Detected-vx	& 704776	\\
Sent-vy	 & 100320000000	   &Detected-vy	& 763880	\\
Sent-vz	 & 455352000000	   &Detected-vz	& 4284050	\\
Sent-xv	 & 308550000000	   &Detected-xv	& 732498	\\
Sent-xx	 & 65028000000	   &Detected-xx	& 289068 \\
Sent-xy	 & 22315000000	   &Detected-xy	& 229251	\\
Sent-xz	 & 98481000000	   &Detected-xz	& 1185899	\\
Sent-yv	 & 104063000000	   &Detected-yv	& 826665	\\
Sent-yx	 & 20424000000     &Detected-yx	& 221359	\\
Sent-yy	 & 7434000000      &Detected-yy	& 117590  \\
Sent-yz	 & 33492000000     &Detected-yz	& 581548  \\
Sent-zv	 & 455417000000    &Detected-zv	& 4411898 \\
Sent-zx	 & 94751000000 	   &Detected-zx	& 1141914 \\
Sent-zy	 & 33479000000     &Detected-zy	& 577524 \\
Sent-zz	 & 150480000000    &Detected-zz	& 2880529 \\
\hline
$n_1$(Before AOPP)		 & 6386090  &$n_1$(After AOPP) & 1099940 \\
$e_1^{ph}$(Before AOPP)  & 6.778\% &$e_1^{ph}$(After AOPP)   & 13.318\%\\
QBER$(Before AOPP)$ 	 & 28.907\%  &QBER$(After AOPP)$ 	 & 0.392\%  \\
Detected-Valid(Before AOPP) &18973852 &Survived-bits(After AOPP)& 2821710\\
\hline
\end{tabular}
\label{Tab:Result3}
\end{table*}

Different accepted phase difference ranges $Ds$ lead to different detection counts and QBERs in $x$ source. We list the QBERs of the $x$ source when Alice and Bob send decoy states $\mu_x$ with different phase difference range ($Ds$), which are listed in Tab.~\ref{Tab:Result1QBERXX}, Tab.~\ref{Tab:Result2QBERXX} and Tab.~\ref{Tab:Result3QBERXX}. It also shows the different detection counts according to different $Ds$ and the optimized secure key rates that we extract by searching through ranges of these parameter values.

\begin{table*}[htb]
\centering
  \caption{QBERs and detections of xx and key rates in 201 km experiment.}
\begin{tabular}{c|cccccc}
\hline
\hline
Results$\mid$Ds/2   & deg=2$^\circ$	& deg=8$^\circ$	& deg=10$^\circ$	& deg=12$^\circ$	& deg=30$^\circ$	& deg=45$^\circ$\\
\hline

QBER(xx)	& 2.9\% & 3.0\% & 3.1\% & 3.4\% & 5.1\% & 7.5\%  \\
Detections of xx & 314164 & 919356  & 1131439 & 1341434 & 3114407 & 4452916 \\
Key Rates & $7.52\times 10^{-5}$ & $8.74\times 10^{-5}$ & $8.69\times 10^{-5}$ & $8.62\times 10^{-5}$ & $8.45\times 10^{-5}$ & $7.08\times 10^{-5}$  \\

\hline
\end{tabular}
\label{Tab:Result1QBERXX}
\end{table*}

\begin{table*}[htb]
\centering
  \caption{QBERs and detections of xx and key rates in 301 km experiment.}
\begin{tabular}{c|cccccc}
\hline
\hline
Results$\mid$Ds/2   & deg=2$^\circ$	& deg=8$^\circ$	& deg=10$^\circ$	& deg=12$^\circ$	& deg=30$^\circ$	& deg=45$^\circ$\\
\hline

QBER(xx)	& 3.1\% & 3.1\% & 3.2\% & 3.3\% & 5.0\% & 7.7\%  \\
Detections of xx & 45014 & 138645  & 171843 & 204812 & 506462 & 766159 \\
Key Rates & $9.63\times 10^{-6}$ & $1.15\times 10^{-5}$ & $1.14\times 10^{-5}$ & $1.13\times 10^{-5}$ & $1.12\times 10^{-5}$ & $9.18\times 10^{-6}$  \\

\hline
\end{tabular}
\label{Tab:Result2QBERXX}
\end{table*}

\begin{table*}[htb]
\centering
  \caption{QBERs and detections of xx and key rates in 502 km experiment.}
\begin{tabular}{c|cccccc}
\hline
\hline
Results$\mid$Ds/2   & deg=2$^\circ$	& deg=8$^\circ$	& deg=10$^\circ$	& deg=12$^\circ$	& deg=30$^\circ$	& deg=45$^\circ$\\
\hline

QBER(xx)	& 4.5\% & 4.4\% & 4.5\% & 4.6\% & 4.9\% & 6.6\%  \\
Detections of xx & 7951 & 27371  & 33939 & 40420 & 50128 & 97855 \\
Key Rates & $8.50\times 10^{-8}$ & $9.65\times 10^{-8}$ & $9.67\times 10^{-8}$ & $9.28\times 10^{-8}$ & $6.98\times 10^{-8}$ & $3.71\times 10^{-8}$  \\

\hline
\end{tabular}
\label{Tab:Result3QBERXX}
\end{table*}

\clearpage
\bibliographystyle{unsrt}
\bibliography{SeismicSNSTFQKD}

\end{document}